\documentclass[epj]{svjour}
\usepackage[utf8]{inputenc}

\usepackage{amsmath}
\usepackage{amssymb}
\usepackage{cite}
\usepackage{graphicx}
\usepackage[normalem]{ulem}

\usepackage{color}

\newcommand{\ie}{\textit{i.e.}}
\newcommand{\eg}{\textit{e.g.}}

\begin{document}

\title{Multimessenger Parameter Estimation of GW170817}
\author{ %
  David Radice\inst{1,2} \and
  Liang Dai\inst{1,3} %
}
\institute{ %
Institute for Advanced Study, 1 Einstein Drive, Princeton, NJ 08540, USA \and
Department of Astrophysical Sciences, Princeton University, 4 Ivy Lane, Princeton, NJ 08544, US \and
NASA Einstein Fellow}
\date{Received: date / Revised version: date}
%
\abstract{
We combine gravitational wave (GW) and electromagnetic (EM) data to perform a Bayesian parameter estimation of the binary neutron star (NS) merger GW170817. The EM likelihood is constructed from a fit to a large number of numerical relativity simulations which we combine with a lower bound on the mass of the remnant's accretion disk inferred from the modeling of the EM light curve. In comparison with previous works, our analysis yields a more precise determination of the tidal deformability of the binary, for which the EM data provide a lower bound, and of the mass ratio of the binary, with the EM data favoring a smaller mass asymmetry. The 90\% credible interval for the areal radius of a $1.4\ M_\odot$ NS is found to be $12.2^{+1.0}_{-0.8} \pm 0.2\ {\rm km}$ (statistical and systematic uncertainties).
} 

\maketitle

\section{Introduction}
\label{sec:intro}

The detection of gravitational waves (GWs) and light from the binary neutron star (NS) merger GW170817 \cite{theligoscientific:2017qsa, abbott:2018wiz, gbm:2017lvd} last year inaugurated the era of multimessenger astronomy with GWs. The electromagnetic (EM) counterpart, now called AT2017gfo/GRB 170817A, had thermal and non-thermal components. The latter consists of a prompt gamma-ray flash generated by a relativistic outflow \cite{monitor:2017mdv} and long lasting synchrotron emission powered by the interaction of this outflow with the interstellar medium (ISM) \cite{kasliwal:2017ngb, margutti:2017cjl, mooley:2017enz, lazzati:2017zsj, margutti:2018xqd, alexander:2018dcl, mooley:2018qfh, ghirlanda:2018uyx}. The thermal component, the so-called kilonova (kN), is thought to have been powered by the radioactive decay of ${\sim}0.03{-}0.06\, M_\odot$ of NS matter ejected during and shortly after the merger \cite{chornock:2017sdf, cowperthwaite:2017dyu, drout:2017ijr, nicholl:2017ahq, tanaka:2017qxj, tanvir:2017pws, perego:2017wtu, villar:2017wcc, waxman:2017sqv, metzger:2018uni, kawaguchi:2018ptg}.

These landmark observations had a far-reaching impact in nuclear and
high-energy astrophysics. The GW data have been used to constrain the NS
tidal deformability \cite{theligoscientific:2017qsa, abbott:2018wiz,
De:2018uhw, abbott:2018exr} and to derive new bounds on the poorly known
equation of state (EOS) of matter at supernuclear densities
\cite{annala:2017llu, fattoyev:2017jql, most:2018hfd, tews:2018iwm,
malik:2018zcf, abbott:2018exr, tsang:2018kqj}. The non-thermal EM
counterpart provided the first direct evidence that NS mergers power
short gamma-ray bursts (sGRBs) \cite{Paczynski:1986px, Eichler:1989ve,
Nakar:2007yr, Berger:2013jza, monitor:2017mdv}, and the thermal
counterpart confirmed that NS mergers are one of the main sites of production of $r$-process elements \cite{kasen:2017sxr, hotokezaka:2018aui}.

The inclusion of sky position and distance information obtained from the EM observations into the GW Bayesian data analysis allowed for a tighter determination of some of the binary parameters \cite{finstad:2018wid, De:2018uhw, abbott:2018wiz}. A joint GW and EM analysis has also been used to measure the Hubble constant \cite{abbott:2017xzu, hotokezaka:2018dfi}. Refs.~\cite{Gao:2017fcu, Pankow:2018iab} proposed to combine EM and GW data to constrain the mass ratio of the two NSs. Moreover, the EM data suggest that the merger resulted neither in prompt black hole (BH) formation, nor in the formation of a long-lived remnant \cite{margalit:2017dij}. This observation has been used to derive additional constraints on the NS EOS and, in particular, on the maximum  mass for a nonrotating NS \cite{margalit:2017dij, shibata:2017xdx, rezzolla:2017aly, ruiz:2017due}. Ref.~\cite{bauswein:2017vtn} used an empirical relation between the threshold mass for prompt BH formation and the radius of the $1.6$-$M_\odot$ NS to place a lower bound on the latter. Ref.~\cite{radice:2017lry} pointed out that the EM observations also imply a lower limit on the tidal parameter $\tilde\Lambda$, \eg, Refs.~\cite{flanagan:2007ix, favata:2013rwa}. This is because, on the one hand, the amount of material ejected during merger is weakly dependent on $\tilde\Lambda$. On the other hand, the overall ejecta and associated kN are dominated by neutrino- and viscous-driven winds from the accretion disk and the mass of the latter strongly depends on $\tilde\Lambda$ \cite{Radice:2018pdn}. A similar approach, but based on the assumption that the outflow was proportional to the amount of the dynamical ejecta, has been proposed by Ref.~\cite{Coughlin:2018miv}.

Here, we extend the work of Ref.~\cite{radice:2017lry}. We incorporate numerical relativity results in a joint Bayesian analysis of the GW and EM data, and we improve the measurement on the binary mass ratio and the tidal deformability. The approach we present here is fully general and will become even more powerful when more accurate simulations spanning a larger portion of the binary parameter space become available.
The remainder of this paper is organized as follows. We discuss the numerical simulations and the setup for the Bayesian analysis in Sec.~\ref{sec:methods}. We give an account of our results in Sec.~\ref{sec:results}. Finally, Sec.~\ref{sec:conclusions} is dedicated to discussion and conclusions.

\section{Methods}
\label{sec:methods}

We perform Bayesian parameter estimation using the combined GW and EM data to determine posteriors for the binary parameters $\theta = \{\mathcal{M}^{\rm det}, q, \chi_{\rm eff}, \chi_a, \tilde\Lambda, t_{c,1}, t_{c,2} \}$, where $\mathcal{M}^{\rm det}= (1+z)\,(M_1\,M_2)^{3/5}/(M_1 + M_2)^{1/5}$ is the detector-frame chirp mass, $q = M_2/M_1 \leq 1$ is the binary mass ratio, $\chi_{\rm eff} = (M_1 \chi_{1z} + M_2 \chi_{2z})/(M_1 + M_2)$ and $\chi_a = (\chi_{1z} - \chi_{2z})/2$ are the parameters describing spin components aligned with the binary orbital angular momentum, and $t_{c,1}$ and $t_{c,2}$ are the arrival times at Livingston and at Hanford, respectively. Not aiming to measure the source's orientation and its sky position, we independently maximize the likelihood at each detector with respect to a constant wave phase and an amplitude normalization, and we assume that $t_{c, 1}$ and $t_{c, 2}$ can be independently adjusted. This approximation greatly simplifies the parameter estimation by reducing the number of parameters. Since GW170817 has a high matched filtering signal-to-noise ratio (SNR), this simplification does not bias the maximum-likelihood values of the parameters but only leads to percent-level increase of their uncertainties~\cite{Roulet:2018jbe}.

Assuming GW and EM data to be independent, we can write the joint GW and EM likelihood as the product of the separate likelihoods, namely
\begin{equation}
	P\big[\{d_{\rm GW}, d_{\rm EM} \} | \theta\big] = P[d_{\rm GW} | \theta] \,P[d_{\rm EM} | \theta],
\end{equation}
where $d_{\rm GW}$ and $d_{\rm EM}$ denote the GW and EM data, respectively.

We compute the first factor with the relative binning method~\cite{zackay:2018qdy, dai:2018dca}. We use the noise-subtracted LIGO data release\footnote{In the noise-substracted data release, the glitch that happened to overlap with GW170817 in the Livingston strain has been removed by the LIGO/Virgo collaboration.} of GW170817 and include frequencies in the range $[23,\,1000]\,$Hz. The exclusion of higher frequency GW data results in a slightly broader posterior of $\tilde\Lambda$ whose support also extends to somewhat larger values, as discussed in detail in Ref.~\cite{dai:2018dca}. It is important, however, to remark that the two NSs first touch when the GW frequency is between 1.0~kHz and 1.5~kHz \cite{Damour:2009wj}. It is thus not clear whether or not current waveform models, which are typically constructed by adding tidal corrections to point particle models, are reliable past 1~kHz, \eg, Ref.~\cite{Kawaguchi:2018gvj}. Consequently, to be conservative, we restrict our analysis to the part of the GW signal below frequency of 1~kHz, which is theoretically well understood. We use the phenomenological waveform model \texttt{IMRPhenomD\_NRTidal} \cite{Dietrich:2017aum, Dietrich:2018uni} implemented in \texttt{LALSuite}.

We follow Ref.~\cite{abbott:2018wiz} for the choice of priors. Both component masses have flat priors in the range $[0.5,\,7.7]\ M_\odot$. The two dimensionless spin vectors have their moduli uniformly distributed in $[0,\,0.89]$ and have isotropic orientations. Their aligned components are then extracted and used to evaluate the non-precessing waveform model \linebreak \texttt{IMRPhenomD\_NRTidal}.

Following the prescription of Ref.~\cite{De:2018uhw}, we relate the component tidal deformability parameters through $\Lambda_1 = \Lambda_s\,q^3$ and $\Lambda_2 = \Lambda_s/q^3$, where $\Lambda_s$ is assigned a uniform prior within $[0,\,5000]$. This implicitly assumes that no first-order phase transition occurs in matter at densities intermediate between those achieved in the secondary and in the primary NS, so that the two NS radii are comparable. Note that the error introduced assuming that the NSs have a commensurate radii is much smaller than current statistical errors \cite{De:2018uhw}. This choice is also consistent with the use of data from our simulations not accounting for the possibility of first order phase transitions in dense matter. Finally, we exclude $\tilde\Lambda > 5000$ which is unreasonable with any plausible EOS.

\begin{figure}
  \includegraphics[width=0.98\columnwidth]{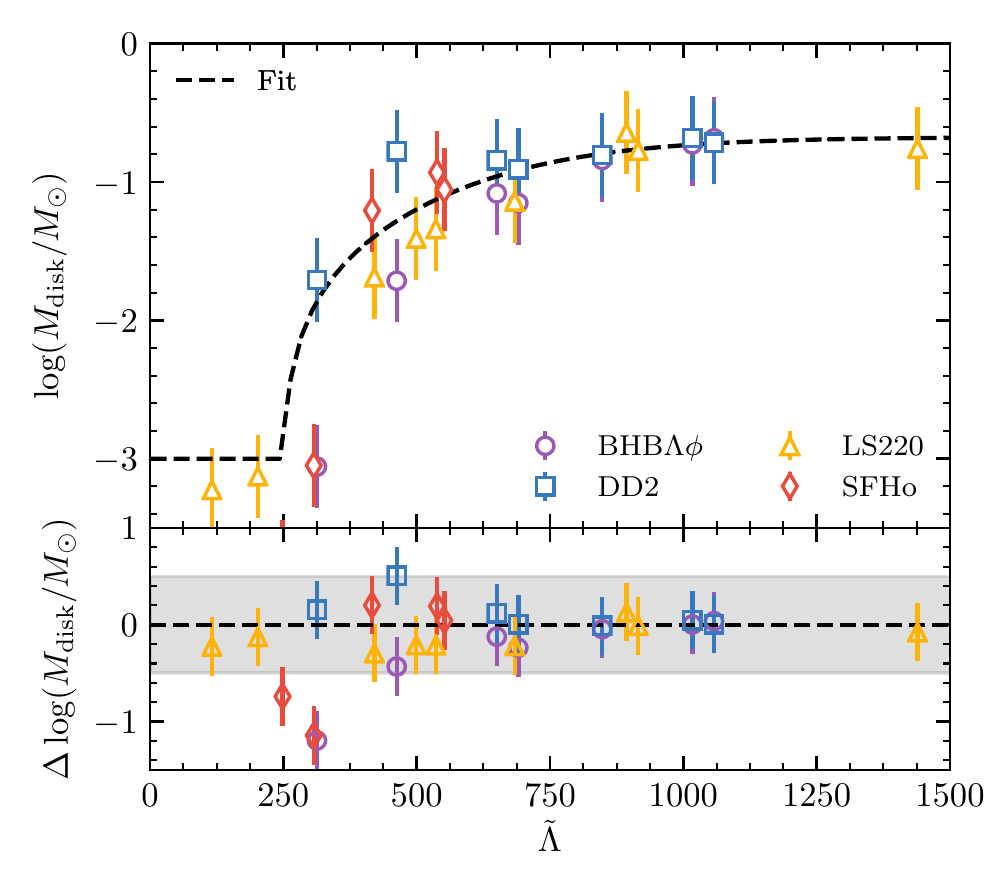}
  \caption{Remnant disk mass as a function of the tidal deformability parameter $\tilde\Lambda$. The data points show the results from our simulations, while the dashed line shows the fit in the form of Eq.~(\ref{eq:mdisk_fit}). The gray shaded region in the lower panel shows the uncertainty $\sigma$ we use in Eq.~(\ref{eq:pdisk}). We find that disk formation is suppressed in the case of prompt BH formation.}
  \label{fig:mdisk}
\end{figure}

Current models of the EM signal are not yet sufficiently advanced to follow the same procedure as for the GW data. However, extant light curve models indicate that $0.02{-}0.05\, M_\odot$ of material with a broad distribution in electron fraction and asymptotic velocity of ${\sim}0.1\, c$ is needed to explain the observations \cite{tanaka:2017qxj, perego:2017wtu, villar:2017wcc, waxman:2017sqv, kawaguchi:2018ptg}. Because of their properties, these ejecta are thought to originate from winds launched from the remnant accretion disk after merger, \eg, Ref.~\cite{Metzger:2017wot}. Long term simulations of postmerger disks indicate that these winds can entrain $10{-}40\, \%$ of the total disk mass \cite{dessart:2008zd, metzger:2008av, metzger:2008jt, lee:2009a, fernandez:2013tya, siegel:2014ita, just:2014fka, metzger:2014ila, perego:2014fma, martin:2015hxa, wu:2016pnw, siegel:2017nub, lippuner:2017bfm, fujibayashi:2017xsz, fujibayashi:2017puw, siegel:2017jug, metzger:2018uni, radice:2018xqa, fernandez:2018kax}. Consequently, we can conservatively estimate that a disk of at least $0.04\, M_\odot$ should have formed in GW170817. Accordingly, we approximate the EM likelihood as
\begin{equation}
	P[d_{\rm EM} | \theta] \simeq P[M_{\rm disk}(\theta) > 0.04\, M_\odot].
\end{equation}

We have performed numerical relativity simulations of merging NS using the \texttt{WhiskyTHC} code \cite{radice:2012cu, radice:2013hxh, radice:2013xpa}.  We considered 29 binaries, including both equal and unequal mass configurations and 4 temperature and composition dependent nuclear EOSs: the DD2 EOS \cite{typel:2009sy, hempel:2009mc}, the BHB$\Lambda\phi$ EOS \cite{banik:2014qja}, the LS220 EOS \cite{lattimer:1991nc}, and the SFHo EOS \cite{steiner:2012rk}. The simulations included temperature and compositional changes due to the emission of neutrinos using a leakage scheme \cite{radice:2016dwd}. A detailed account of the numerical results is given in Refs.~\cite{radice:2017lry, radice:2018xqa, Radice:2018pdn}.

The simulation data suggest that the remnant disk masses can be related to the tidal deformability parameter $\tilde\Lambda$ through the fitting formula \cite{Radice:2018pdn}
\begin{equation}\label{eq:mdisk_fit}
\begin{split}
  &\log\left(\frac{M_{\rm disk}}{M_\odot}\right) \simeq \Phi(\tilde\Lambda) := \\ 
  &\qquad\qquad \max \left\{ -3, \log\left[\alpha + \beta \tanh\left(
  \frac{\tilde\Lambda - \gamma}{\delta} \right)\right] \right\},
\end{split}
\end{equation}
with coefficients $\alpha = 0.084$, $\beta = 0.127$, $\gamma = 567.1$, and $\delta = 405.14$. The numerical data, the best fit, and the residual are shown in Fig.~\ref{fig:mdisk}. We remark that our simulations have only sampled the region of parameter space with $q \geq 0.85$. Smaller mass ratios could result in larger disk masses for a given $\tilde\Lambda$. However, the variation of $M_{\rm disk}$ with $q$ reported in the literature, \eg, Refs.\cite{Shibata:2006nm, Rezzolla:2010fd}, are not large enough to affect our results in a qualitative way. Moreover, large mass asymmetries are disfavored in the light of the distribution of known binary NS systems in our galaxy \cite{Tauris:2017omb}. We leave the determination of $M_{\rm disk}$ as a function of $q$ to future work.

For the likelihood calculation we assume $\log (M_{\rm disk}/M_\odot)$ to have a Gaussian distribution with mean $\Phi(\tilde\Lambda)$. We conservatively take the standard deviation to be $\sigma=0.5$. This uncertainty is indicated by the gray shaded region in the bottom panel of Fig.~\ref{fig:mdisk}. Accordingly, we approximate the EM likelihood as
\begin{equation}\label{eq:pdisk}
\begin{split}
	P[M_{\rm disk} > & 0.04\, M_\odot] = \\
    &1 - \frac{1}{2}\left[ 1 + \mathrm{erf}\left(\frac{\log(0.04) - \Phi(\tilde\Lambda)}{\sqrt{2} \sigma}\right) \right].
\end{split}
\end{equation}
To explore the parameter space and obtain posterior samples, we couple the evaluaton of the likelihood function to \texttt{MultiNest}~\cite{Feroz:2008xx}. This is a Monte Carlo sampling algorithm that uses the technique of nested sampling, and is designed to efficiently cope with disjoint multi-modal posteriors in the multi-dimensional parameter space.

\section{Results}
\label{sec:results}

\begin{figure*}
	\includegraphics[width=\textwidth]{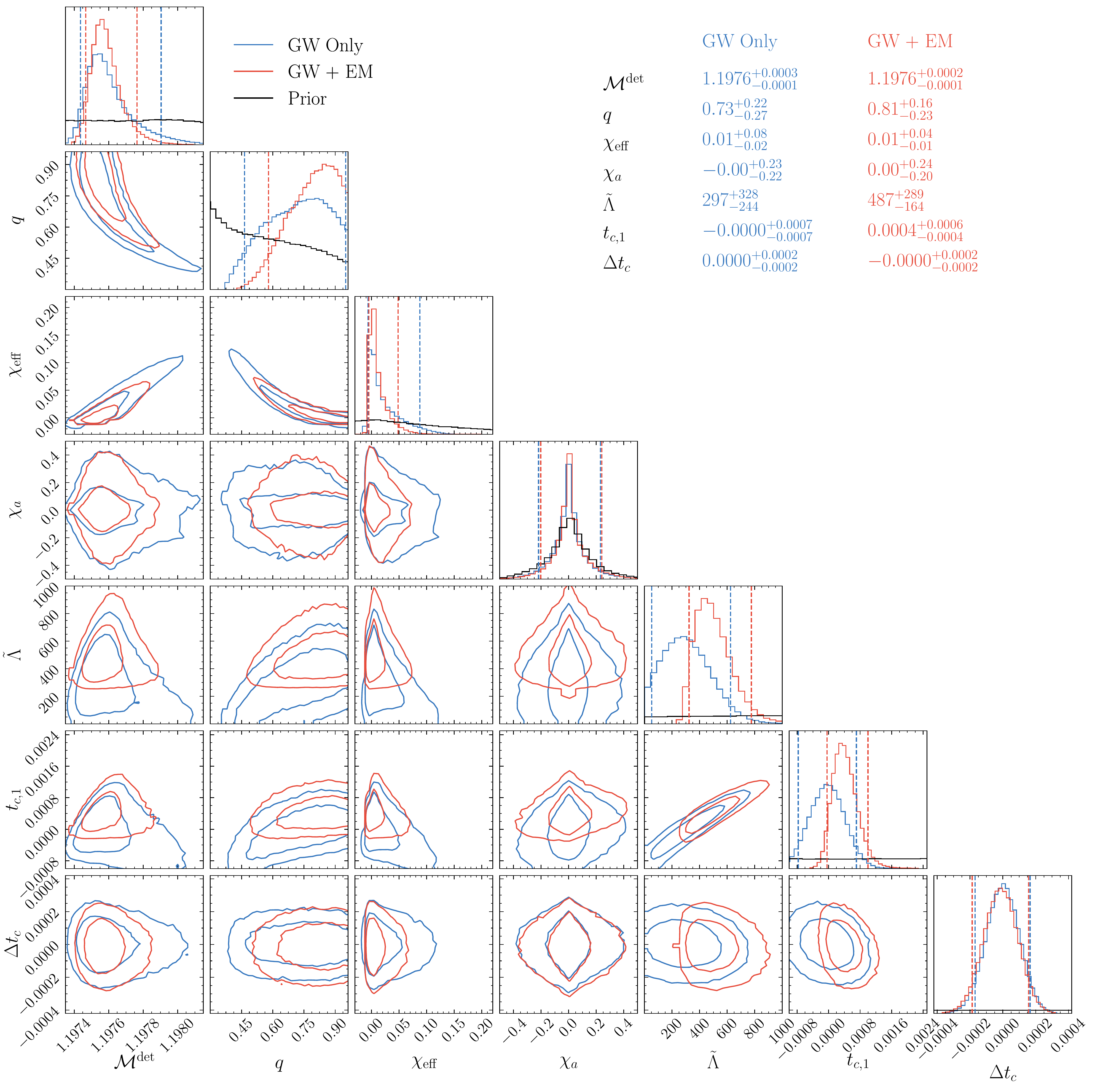}
    \caption{Posterior distributions obtained with (red) and without (blue) the inclusion of the EM constraints. The marginalized prior distribution for each parameter is shown as the black histogram in the plots along the diagonal. The off-diagonal plots show contours enclosing 68\% and 95\% quantiles for the two-dimensional joint posterior distributions. On the upper right corner, we indicate for each of the parameters the median value and the uncertainty (also shown by the vertical lines in the plots along the diagonal). The uncertainty corresponds to the 5\% and 95\% percentiles. Instead of showing the two arrival times $t_{c, 1}$ and $t_{c, 2}$ separately, we show $t_{c, 1}$ (Livingston) and $\Delta t_c = t_{c, 2} - t_{c, 1}$. Our chosen zero point for $t_{c, 1}$ is $0.0035\,$s in advance of that for $t_{c, 2}$. The results are consistent with the causality bound on the time delay between the two LIGO sites. The EM data favors larger values of the tidal deformability parameter $\tilde\Lambda$ and of the mass ratio $q$, \ie, larger NS radii and more symmetric binaries.}
    \label{fig:posteriors}
\end{figure*}

The results of our analysis are summarized in Fig.~\ref{fig:posteriors}. There we show the marginalized 1-parameter histograms as well as the marginalized 2-parameter joint distributions for the posterior samples obtained both with and without including the EM data in the likelihood. The results clearly show that the mutual delay $\Delta t_c := t_{c, 2} - t_{c, 1}$ between the two LIGO detectors does not correlate with any of the intrinsic parameters, and that the independently inferred arrival times at the two sites do not differ by more than the causality bound. These justify the simplification that we have ignored the time, phase and amplitude correlations between the GW signals recorded at both detectors.

Our GW-only posteriors are consistent with those presented in Refs.~\cite{theligoscientific:2017qsa, abbott:2018wiz, De:2018uhw, abbott:2018exr}.See \cite{dai:2018dca} for a more detailed discussion of the GW-only posteriors obtained with our approach. However, our posterior for $\tilde\Lambda$ is broader because of the  more conservative choice of cutoff frequency for the GW data \cite{dai:2018dca}. This is expected because $\tilde\Lambda$ is mostly encoded in the high-frequency part of the GW signal \cite{Damour:2009wj, De:2018uhw}. Also note that there is a degeneracy between $\tilde\Lambda$ and the {\it common} arrival time of the two detectors. This is because both tidal deformability and the arrival time cause phasing corrections that grow as positive powers of the frequency $f$, with similar power indices: $5/3$ and $1$, respectively \cite{dai:2018dca}.

\begin{figure}
  \includegraphics[width=0.98\columnwidth]{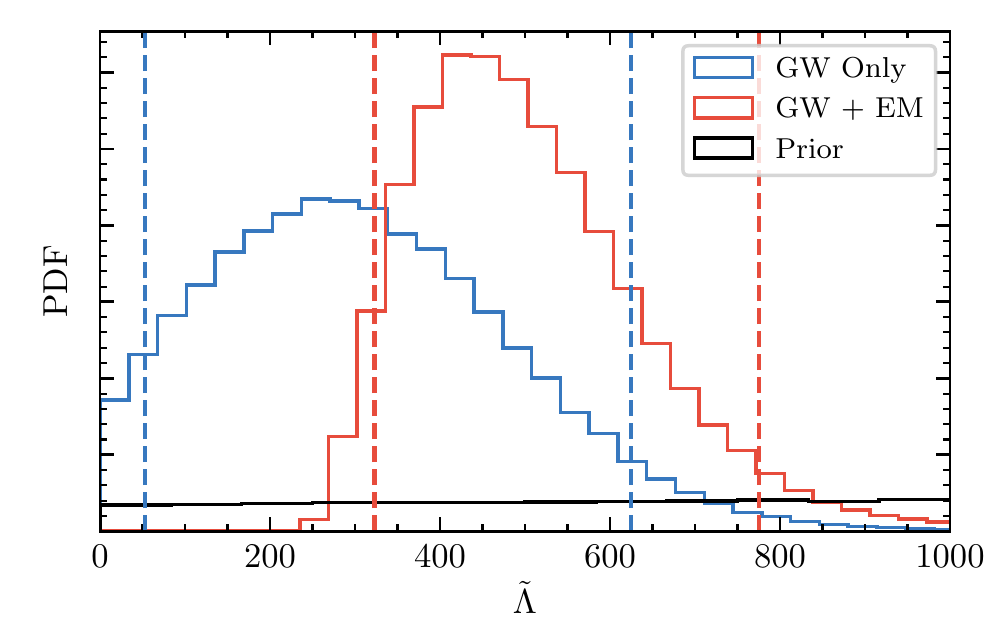}
  \caption{Posterior distribution function for $\tilde\Lambda$ obtained with (red) and without (blue) the inclusion of the EM constraints. The inclusion of EM information shifts the posterior towards larger values of $\tilde\Lambda$ and further away from zero.}
  \label{fig:lambda}
\end{figure}

The inclusion of EM information in the likelihood function has a strong impact on the recovered posterior for $\tilde\Lambda$, shown in Fig.~\ref{fig:lambda}. Values of $\tilde\Lambda$ smaller than about $300$ appear to be excluded by the EM data and the overall distribution for $\tilde\Lambda$ is shifted towards larger values. The 90\% confidence interval for $\tilde\Lambda$ shifts from $(53, 625)$ with median $297$ to $(323, 776)$ with median $487$. Other parameters that correlate with $\tilde\Lambda$ are also affected. The most notable is the binary mass ratio $q$, with the EM data favoring more comparable component masses (see Fig.~\ref{fig:posteriors}).

The lower limit $\tilde\Lambda \gtrsim 300$ is not as stringent as the one in Ref.~\cite{radice:2017lry}, who found $\tilde\Lambda \gtrsim 400$. The reason for this discrepancy is that, in the analysis performed here, the probability of forming an accretion disk with a mass more than one standard deviation larger than $\Phi(\tilde\Lambda)$ is not zero, while it was instead implicitly assumed so in Ref.~\cite{radice:2017lry}. On the other hand, we want to emphasize that the goal of Ref.~\cite{radice:2017lry} was not to perform a fully quantitative analysis, as we have done here, but only to illustrate the key idea. In this sense, our results and those of Ref.~\cite{radice:2017lry} are fully consistent.

We can translate the measurement of $\tilde\Lambda$ into a constraint on the radius of a $1.4\ M_\odot$ NS following Refs.~\cite{De:2018uhw, Zhao:2018nyf}. They derived the EOS insensitive relation
\begin{equation}\label{eq:r14}
	R_{14} = (11.2 \pm 0.2) \frac{\mathcal{M}}{M_\odot} \left( \frac{\tilde\Lambda}{800} \right)^{1/6}\ {\rm km},
\end{equation}
To apply this formula we compute the rest-frame binary chirp mass from the detector-frame chirp mass as $\mathcal{M} (1 + z) = \mathcal{M}^{\rm det}$, where $z$ is taken to be $0.0099$ following Ref.~\cite{abbott:2018wiz}. Accordingly, we find the median value of $\mathcal{M}$ to be $1.186\ M_\odot$. From the GW data alone we infer $R_{14} = (11.3^{+1.5}_{-2.8} \pm 0.2)\ {\rm km}$ (90\% credible interval, statistical and systematic uncertainties). With the additional constraint due to the EM data we find $R_{14} = (12.2^{+1.0}_{-0.8} \pm 0.2)\ {\rm km}$. The systematic errors in this estimate include only the uncertainty related to the use of Eq.~(\ref{eq:r14}), but not the possible systematic effects in our numerical relativity data, which we cannot presently quantify. Notwithstanding this caveat, our estimates provide the tightest constraint on the NS radius to date, with an uncertainty of only $2.2~{\rm km}$. Moreover, our analysis strongly disfavors NS radii smaller than $11.2\ {\rm km}$, which would have resulted in early BH formation and would have created accretion disks not sufficiently massive to fuel the outflow inferred from the kN observations.

\section{Conclusions}
\label{sec:conclusions}

We have performed a Bayesian parameter estimation analysis of GW170817/AT2017gfo combining both the GW and the EM data. Specifically, we have argued that EM observations imply a lower limit on the merger remnant disk mass of $0.04\ M_\odot$, and we have used a fit to the simulation data to estimate the probability with which such constraint is fulfilled depending on the binary tidal deformability parameter $\tilde\Lambda$. Then, we have assumed GW and EM data to be independent, and we have employed this probability to construct a joint likelihood for the GW and the EM data. We have used the relative binning method to efficiently evaluate the GW part of the likelihood, while the EM part of the likelihood is analytic. Finally, we have derived the posterior probabilities for binary parameters using a multimodal nested sampler.

We find that the inclusion of the EM information shifts the support of the posterior distribution for $\tilde\Lambda$ to larger values than those inferred from the GW data alone. In particular, values of $\tilde\Lambda$ less than ${\sim}300$ are excluded. This corresponds to a lower limit on the radius of a $1.4\ M_\odot$ NS $R_{14}$ of $11.2\ {\rm km}$. The 90\% credible interval for $R_{14}$ is found to be $12.2^{+1.0}_{-0.8}\ {\rm km}$ with an additional $0.2\ {\rm km}$ of systematic uncertainty. EM data also favors larger values of $q$, \ie, a more symmetric binary, compared to inference from the GW data alone.

We have assumed that both NSs in GW170817 had  similar radii, following \cite{De:2018uhw}. However, this hypothesis would be violated in the presence of first order phase transition at densities intermediate between those achieved in the primary and in the secondary NS. Such scenario, the so-called twin star hypothesis, is presently not excluded for GW170817 \cite{Paschalidis:2017qmb}. If GW170817 was the merger of a regular NS with an hybrid star or a quark star, then our analysis would be invalid. The empirical formula used to relate $\Lambda_1$ and $\Lambda_2$ and Eq.~(\ref{eq:r14}) can be extended to deal with phase transitions, but only at the price of significantly larger systematic errors \cite{Zhao:2018nyf}. Perhaps more importantly, our analysis relies on fits to a relatively large, but still limited set of numerical relativity simulations that do not include examples with first order phase transitions. Additional simulations, spanning a larger range of the parameter space and more EOSs and including cases with first-order phase transitions, would be required to confirm our results. This will be the object of our future work.

\subsection*{Acknowledgments}

It is a pleasure to acknowledge Albino Perego, Sebastiano Bernuzzi, Tim Dietrich, Ingo Tews, Sanjay Reddy, Matias Zaldarriaga, and Adam Burrows for discussions.
DR acknowledges support from a Frank and Peggy Taplin Membership at the Institute for Advanced Study and the Max-Planck/Princeton Center (MPPC) for Plasma Physics (NSF PHY-1804048).
Computations were performed on the supercomputers Bridges, Comet, and Stampede (NSF XSEDE allocation TG-PHY160025), on NSF/NCSA Blue Waters (NSF PRAC ACI-1440083 and AWD-1811236), and on CINECA's Marconi (PRACE proposal 2016153522). LD is partially supported at the Institute for Advanced Study by NASA through Einstein Postdoctoral Fellowship grant number PF5-160135 awarded by the Chandra X-ray Center, which is operated by the Smithsonian Astrophysical Observatory for NASA under contract NAS8-03060. LD is also supported at the Institute for Advanced Study by the Raymond and Beverly Sackler Foundation.

\bibliographystyle{epj}
\bibliography{references,local}

\end{document}